\documentclass[english,aps, reprint, prb, amsmath,amssymb,floatfix,longbibliography]{revtex4-2}
\usepackage[T1]{fontenc}
\usepackage[latin9]{inputenc}
\usepackage{babel}
\usepackage{amsmath}
\usepackage{amssymb}
\usepackage{graphicx}
\usepackage[unicode=true,pdfusetitle,
 bookmarks=true,bookmarksnumbered=false,bookmarksopen=false,
 breaklinks=false,pdfborder={0 0 1},backref=false,colorlinks=false]
 {hyperref}

\makeatletter
\usepackage{braket}

\makeatother

\begin{document}
\title{Difference between angular momentum and pseudoangular momentum}
\author{Simon Streib}
\affiliation{Department of Physics and Astronomy, Uppsala University, Box 516,
SE-75120 Uppsala, Sweden}
\date{March 16, 2021}
\begin{abstract}
In condensed matter systems it is necessary to distinguish between
the momentum of the constituents of the system and the pseudomomentum
of quasiparticles. The same distinction is also valid for angular
momentum and pseudoangular momentum. Based on Noether's theorem,
we demonstrate that the recently discussed orbital angular momenta
of phonons and magnons are pseudoangular momenta. This conceptual
difference is important for a proper understanding of the transfer
of angular momentum in condensed matter systems, especially in spintronics
applications. 
\end{abstract}
\maketitle
In 1915, Einstein, de Haas, and Barnett demonstrated experimentally
that magnetism is fundamentally related to angular momentum. When
changing the magnetization of a magnet, Einstein and de Haas observed
that the magnet starts to rotate, implying a transfer of angular momentum
from the magnetization to the global rotation of the lattice \cite{Einstein15},
while Barnett observed the inverse effect, magnetization by rotation
\cite{Barnett1915}. A few years later in 1918, Emmy Noether showed
that continuous symmetries imply conservation laws \cite{Noether1918},
such as the conservation of momentum and angular momentum, which links
magnetism to the most fundamental symmetries of nature. 

Condensed matter systems support closely related conservation laws:
the conservation of the pseudomomentum and pseudoangular momentum
of quasiparticles, such as magnons and phonons. While the distinction
between momentum and pseudomomentum (or quasimomentum) is well-known
\cite{Gordon1973,Peierls1976,Peierls1979,Thellung1980,Mcintyre1981,Nelson1991,Peierls1991,Thellung1994,Stone2002,Heller2019}
and is crucial for understanding the momentum of light in dielectric
materials \cite{Pfeifer2007}, the concept of pseudoangular momentum
has not been widely discussed in the literature, so far only in the
context of optical and acoustical vortices \cite{Thomas2003} and
for chiral phonons with a discrete rotational symmetry \cite{Zhang2015,Zhu2018,Tatsumi2018,ZhangW2020}. 

The transfer of angular momentum between lattice and magnetization
has been investigated in several recent works, for example the dynamics
of a single spin embedded in an elastic solid \cite{Garanin2015,Nakane2018,Mentink2019}
and the transfer of angular momentum in magnetic insulators \cite{Streib2018,Rueckiegel2020,Rueckriegel2020b,Zhang2020,Juraschek2020}.
The spin angular momentum of phonons was first introduced by Vonsovskii
and Svirskii \cite{Vonsovskii1962}, which was later rediscovered
\cite{Zhang2014}, as well as experimentally confirmed \cite{Holanda2018}.
It has been suggested that phonon spin could be used to transfer angular
momentum over longer distances than would be possible with electrons
or magnons \cite{An2020,Rueckriegel2020b,Brataas2020}.

Several recent publications have discussed the orbital angular momentum
of phonons \cite{Ayub2011,Nakane2018,Mentink2019} and magnons \cite{Jia2019,Jiang2019}.
The phonon orbital angular momentum plays an important role in the
relaxation process of a single spin \cite{Nakane2018,Mentink2019},
whereas the magnon orbital angular momentum could be used for the
transfer of information due to its topological stability \cite{Jia2019}
and for the manipulation of Skyrmions \cite{Jiang2019}, analogous
to using the orbital angular momentum of light \cite{Yang2018}. However,
there is a conceptual issue related to these recent advances: neither
phonons nor magnons carry an orbital angular momentum. We demonstrate
in this Letter that these orbital angular momenta are actually pseudo-orbital angular momenta, which phonons and magnons may carry. 

\begin{figure}
\begin{centering}
\includegraphics[scale=1.2]{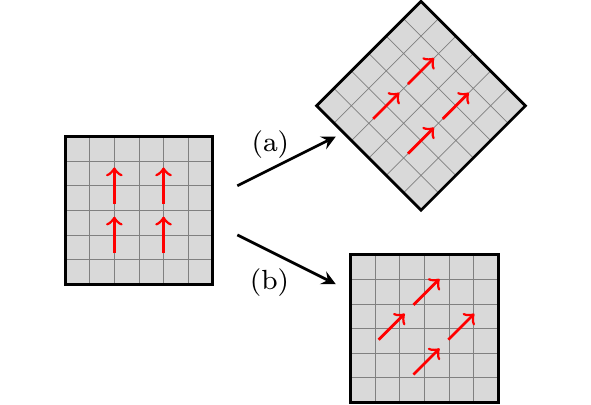}
\par\end{centering}
\caption{(a) Invariance under rotations of the whole system implies conservation
of angular momentum, while (b) invariance under rotations of fields
with a fixed background implies conservation of pseudoangular momentum.\label{fig:pam}}
\end{figure}

The difference between pseudomomentum and momentum can most easily
be explained by considering under which conditions they are conserved.
Conservation of momentum follows from the uniformity of space and
the invariance under translations of the whole system, 
\begin{equation}
\mathbf{r}\to\mathbf{r}+\mathbf{a},\label{eq:translation}
\end{equation}
where $\mathbf{r}$ is the position coordinate and $\mathbf{a}$ an
arbitrary translation vector. The conservation of the pseudomomentum
associated with a field $f(\mathbf{r})$ follows from invariance of
the Lagrangian under translations of the field,
\begin{equation}
f(\mathbf{r})\to f(\mathbf{r}-\mathbf{a}),\label{eq:field translation}
\end{equation}
which is only valid in a uniform medium. For a crystal, pseudomomentum
(``crystal momentum'') is only conserved modulo $\hbar\times\text{reciprocal\,lattice\,vector}$
\cite{Ashcroft}. The pseudomomentum of a quasiparticle with wavevector
$\mathbf{k}$ is simply given by $\hbar\mathbf{k}$, where $\mathbf{k}$
is defined by an expansion of the field in plane waves,
\begin{equation}
f_{\mathbf{k}}(\mathbf{r})\sim e^{i\mathbf{k}\cdot\mathbf{r}}.\label{eq:plane wave}
\end{equation}

The difference between pseudomomentum and momentum arises only for
fields in a medium and not in vacuum where both momenta are the same
since the two transformations (\ref{eq:translation}) and (\ref{eq:field translation})
are then equivalent \cite{Peierls1979,Peierls1991}. The pseudomomentum
of quasiparticles in condensed matter systems is an important quantity
and the qualifier ``pseudo'' is usually omitted in the literature.
A well-known example is the momentum of phonons \cite{Peierls1979,Peierls1991,Kittel}.
The momentum of a phonon with pseudomomentum $\hbar\mathbf{k}$ is
exactly zero since the sum of the momenta of the atoms cancels. We
note that it is possible to define phonons in liquids in such a way
that they carry momentum \cite{Peierls1979,Thellung1980,Thellung1994}.

The definition of pseudoangular momentum is completely analogous
to that of pseudomomentum. While the conservation of angular momentum
follows from invariance under rotations of all the constituents of
a solid, the conservation of pseudoangular momentum follows from
invariance under rotations of fields while keeping the solid fixed,
as illustrated in Fig.~\ref{fig:pam}. The corresponding conserved
quantities can then be derived from Noether's theorem \cite{Noether1918}.
Scalar fields only have an orbital angular momentum, while vector
and spinor fields also have a spin angular momentum, which can be
identified as the intrinsic contribution to the angular momentum that
is independent of the origin of the coordinate system \cite{Soper,Greiner}.
Similarly to pseudomomentum, pseudoangular momentum cannot be exactly
conserved in a real material due to the discrete nature of atoms.
However, for long-wavelength excitations the discrete atomic structure
of a material is not important and the pseudoangular momentum can
be approximately conserved if the material is very isotropic. A prime
example is the magnetic insulator yttrium iron garnet (YIG), for which
the energy dispersion of long-wavelength magnons and phonons is isotropic
\cite{Cherepanov1993,Gurevich1996,Maehrlein2016,Maehrlein2017}. YIG
is widely used in magnon spintronics applications due to its record-low
magnetization damping \cite{Kruglyak2010,Chumak2015,Nikitov2015,Brataas2020}.

While in the case of pseudomomentum we can identify the pseudomomentum
with the wavevector $\mathbf{k}$ of a plane wave, in the case of
pseudoangular momentum we can identify the pseudoangular momentum
projection along the $z$ axis with the angular momentum quantum number
$m$ of an expansion of the field in terms of spherical harmonics
$Y_{lm}$ for spherical geometries \cite{Mentink2019} or in terms
of Bessel functions for cylindrical geometries \cite{Jia2019}. In
both cases the fields have the typical azimuthal angle dependence
\begin{equation}
f_{m}(\mathbf{r})\sim e^{im\phi},
\end{equation}
which is the angular analogue to the plane wave (\ref{eq:plane wave}).

Since we have these two different types of symmetries in condensed
matter, invariance under translations or rotations of the whole system
or only of some fields, we have to consider two different types of
conservation laws: the conservation of momentum and angular momentum
and the conservation of pseudomomentum and pseudoangular momentum.
The elementary particles that constitute condensed matter systems
are photons, electrons, and nuclei, disregarding the decomposition
of nuclei into proton and neutrons or quarks, which is only relevant
at much higher energy scales. For the elementary particles in vacuum
their momentum and pseudomomentum and their angular momentum and pseudoangular momentum are identical, respectively.

The total angular momentum of a material can be decomposed into
\begin{equation}
\mathbf{J}=\mathbf{L}_{e}+\mathbf{S}_{e}+\mathbf{L}_{n}+\mathbf{S}_{n},
\end{equation}
where $\mathbf{L}_{e}$ and $\mathbf{L}_{n}$ are the orbital angular
momenta of the electrons and nuclei and $\mathbf{S}_{e}$ and $\mathbf{S}_{n}$
their spin angular momenta. Both the orbital and spin angular momenta
of the corresponding elementary fields are related to a circulating
momentum flow \cite{Ohanian1986,Sebens2019,Bliokh2020}. In the following,
we will denote the pseudomomentum and pseudoangular momentum by a
tilde, e.g., $\tilde{\mathbf{P}}$ for the pseudomomentum, to distinguish
them from momentum and angular momentum.

As a first example, we consider the inelastic neutron scattering process
where a phonon is created \cite{Peierls1979}. The incoming neutron
carries a momentum and pseudomomentum $\hbar\mathbf{k}$. In the scattering
process the pseudomomentum is transferred to the phonon with pseudomomentum
$\hbar\mathbf{k}$. Since the phonon does not carry momentum, the
momentum $\hbar\mathbf{k}$ of the neutron has to be absorbed by the
center of mass motion of the lattice. As the associated kinetic energy
of a body of mass $M$ is given by $\left(\hbar\mathbf{k}\right)^{2}/(2M)$,
the contribution of this kinetic energy term can be disregarded in
the conservation of energy for macroscopic bodies and the energy lost
by the neutron is completely transferred to the phonon. For a complete
understanding of this process we therefore have to consider both the
conservation of momentum and pseudomomentum. 

When considering symmetries and their corresponding conservation laws,
it is important to take the boundary conditions into account. For
lattice vibrations, periodic boundary conditions imply that the lattice
cannot rotate and angular momentum is not conserved, as it can leave
the system via the boundary conditions. Lenstra and Mandel showed
that the angular momentum of the quantized electromagnetic field is
also not conserved under periodic boundary conditions \cite{Lenstra1982}.
Experimentally, a sample is usually fixed to a sample holder, which
prevents the sample from moving or rotating. Therefore, momentum and
angular momentum are not conserved within the sample, whereas pseudomomentum
and pseudoangular momentum can still be conserved. This has two implications:
the conservation of pseudomomentum and pseudoangular momentum is
immensely useful for describing experimentally relevant situations
and, if we find that the total momentum or angular momentum is not
conserved, it most likely has been lost to the boundary conditions
via the lattice.

Nakane and Kohno have derived from Noether's theorem two different
expression for the angular momentum of phonons \cite{Nakane2018},
which they call angular momentum in field theory and angular momentum
in Newtonian mechanics. Their angular momentum in field theory\emph{
}follows from invariance under global rotations of the phonon (or
displacement) field $\mathbf{u}(\mathbf{r})$,
\begin{equation}
\mathbf{u}(\mathbf{r})\to\mathcal{R}\mathbf{u}\left(\mathcal{R}^{-1}\mathbf{r}\right),
\end{equation}
where the rotation matrix $\mathcal{R}$ appears twice because both
the positions and directions of the fields are rotated. The phonon
field $\mathbf{u}(\mathbf{r})$ is a vector field that describes the
displacement of the atoms at position $\mathbf{r}$. From our discussion
of pseudoangular momentum, it is clear that this angular momentum
corresponds to the pseudoangular momentum of the phonon field, since
it is based on the invariance under rotations of an emergent field.
It takes the form \cite{Nakane2018}
\begin{align}
\tilde{\mathbf{J}}_{\text{ph}} & =\tilde{\mathbf{L}}_{\text{ph}}+\tilde{\mathbf{S}}_{\text{ph}},\\
\tilde{\mathbf{L}}_{\text{ph}} & =\sum_{\alpha=x,y,z}\int\rho\dot{u}_{\alpha}\mathbf{r}\times\left(-\boldsymbol{\nabla}\right)u_{\alpha}d^{3}r,\\
\tilde{\mathbf{S}}_{\text{ph}} & =\int\rho\left(\mathbf{u}\times\dot{\mathbf{u}}\right)d^{3}r,
\end{align}
where $\rho$ is the mass density of the solid and both the orbital
contribution $\tilde{\mathbf{L}}_{\text{ph}}$ and the spin contribution
$\tilde{\mathbf{S}}_{\text{ph}}$ are bilinear in the phonon field.
The corresponding phonon pseudomomentum reads \cite{Nakane2018}
\begin{equation}
\tilde{\mathbf{P}}_{\text{ph}}=\sum_{\alpha=x,y,z}\int\rho\dot{u}_{\alpha}\left(-\boldsymbol{\nabla}\right)u_{\alpha}d^{3}r.
\end{equation}

The ``Newtonian'' angular momentum derives from invariance under
rotations of the whole lattice \cite{Nakane2018},
\begin{equation}
\mathbf{u}(\mathbf{r})+\mathbf{r}\to\mathcal{R}\left[\mathbf{u}(\mathbf{r})+\mathbf{r}\right].
\end{equation}
Therefore this angular momentum is the orbital angular momentum of
the atoms and is given by \cite{Garanin2015,Nakane2018}
\begin{align}
\mathbf{J}_{\text{ph}} & =\mathbf{L}_{\text{ph}}+\mathbf{S}_{\text{ph}}=\int\rho\left(\left[\mathbf{u}+\mathbf{r}\right]\times\dot{\mathbf{u}}\right)d^{3}r,\\
\mathbf{L}_{\text{ph}} & =\int\rho\left(\mathbf{r}\times\dot{\mathbf{u}}\right)d^{3}r,\\
\mathbf{S}_{\text{ph}} & =\int\rho\left(\mathbf{u}\times\dot{\mathbf{u}}\right)d^{3}r,
\end{align}
where the spin contribution $\mathbf{S}_{\text{ph}}$ is formally
identified as the intrinsic contribution which is independent of the
origin of the coordinate system \cite{Garanin2015}. The momentum
of the lattice,
\begin{equation}
\mathbf{P}_{\text{ph}}=\int\rho\dot{\mathbf{u}}\,d^{3}r,
\end{equation}
is only finite when there is a center of mass motion of the whole
lattice, contrary to the pseudomomentum which is finite for phonon
excitations without a center of mass motion.

Comparing the angular momentum and pseudoangular momentum associated
with the phonon field, we see that spin and pseudospin are identical,
while orbital and pseudo-orbital angular momentum are not. Both the
momentum $\mathbf{P}_{\text{ph}}$ and orbital angular momentum $\mathbf{L}_{\text{ph}}$
are linear in the phonon field and vanish for phonon excitations,
e.g., for any plane wave with $\mathbf{k}\neq0$,
\begin{equation}
\mathbf{P}_{\text{ph}}\sim\int e^{i\mathbf{k}\cdot\mathbf{r}}\,d^{3}r=0.
\end{equation}
A finite value of $\mathbf{L}_{\text{ph}}$ corresponds to a global
rotation of the whole lattice. Like $\mathbf{P}_{\text{ph}}$ and
$\tilde{\mathbf{P}}_{\text{ph}}$, $\mathbf{L}_{\text{ph}}$ and $\tilde{\mathbf{L}}_{\text{ph}}$
do not represent the same physical quantity, contrary to the suggestion
in Ref.~\cite{Nakane2018}. The phonon spin is finite for circularly
polarized phonons, analogous to the spin angular momentum of light
\cite{Enk1994}. 

The importance of distinguishing between the angular momentum and
pseudoangular momentum of phonons can be demonstrated by considering
two recent publications. Ref.~\cite{Mentink2019} considers a single
spin embedded in an elastic medium, where angular momentum is transferred
from the spin to the orbital angular momentum of phonons. Since we
have established that phonons do not carry orbital angular momentum,
the phonon orbital angular momentum in Ref.~\cite{Mentink2019} has
to be the pseudo-orbital angular momentum $\tilde{\mathbf{L}}_{\text{ph}}$,
which phonons may carry. This can be confirmed by considering that
the orbital angular momentum of the phonon modes in Ref.~\cite{Mentink2019}
is based on an expansion in spherical harmonics $Y_{lm}$, which is
analogous to an expansion in plane waves with pseudomomentum $\mathbf{k}$.
The angular momentum of the embedded spin enters the conversation
of pseudoangular momentum because the system is only invariant under
rotations of both the phonon fields and the spin direction as both
are coupled \cite{Nakane2018}. Therefore a change in spin has to
be balanced by a change of the phonon pseudoangular momentum. The
key question is: what happens to the angular momentum of the spin
if it is not transferred to the phonons? As Ref.~\cite{Mentink2019}
applies periodic boundary conditions to the phonons, the total angular
momentum is not conserved. Physically, the angular momentum is in
this case absorbed by the boundary conditions and would be transferred
to a global rotation of the lattice if it were not for the boundary
conditions that prevent such a rotation. Ref.~\cite{Rueckiegel2020},
on the other hand, considers the transfer of angular momentum between
magnons, phonons, and the global lattice rotation of a magnetic insulator.
There, the orbital angular momentum of phonons is not mentioned at
all since the quantity under consideration is the angular momentum
of the phonons, which is carried only by the phonon spin. The apparent
contradiction between Refs. \cite{Mentink2019} and \cite{Rueckiegel2020}
is resolved by taking into account that Ref.~\cite{Mentink2019}
is considering the pseudoangular momentum and Ref.~\cite{Rueckiegel2020}
the angular momentum of phonons. Both approaches are valid, but it
is important to be aware of which kind of angular momentum is described
and which conservation laws apply.

Similarly to the phonon case, we analyze next the angular momentum
of magnons in a collinear ferromagnet, motivated by recent advances
related to their orbital angular momentum \cite{Jia2019,Jiang2019}.
We define the classical magnon field $\psi(\mathbf{r})$ in the following
way. We start from the Holstein-Primakoff transformation of the spin
operator $\hat{S}_{i}^{\pm}=\hat{S}_{i}^{x}\pm i\hat{S}_{i}^{y}$
at lattice site $i$ \cite{Holstein40},
\begin{align}
\hat{S}_{i}^{+} & =\hbar\sqrt{2S}\sqrt{1-\frac{\hat{n}_{i}}{2S}}\hat{b}_{i},\\
\hat{S}_{i}^{-} & =\hbar\sqrt{2S}\hat{b}_{i}^{\dagger}\sqrt{1-\frac{\hat{n}_{i}}{2S}},\\
\hat{S}_{i}^{z} & =\hbar\left(S-\hat{n}_{i}\right),
\end{align}
where $S$ is the dimensionless spin quantum number, $\hat{n}_{i}=\hat{b}_{i}^{\dagger}\hat{b}_{i}$
the magnon number operator, and $\hat{b}_{i}^{\dagger}$ and $\hat{b}_{i}$
the magnon creation and annihilation operators with the commutation
relations
\begin{equation}
\left[\hat{b}_{i},\hat{b}_{j}^{\dagger}\right]=\delta_{ij},\quad\left[\hat{b}_{i},\hat{b}_{j}\right]=\left[\hat{b}_{i}^{\dagger},\hat{b}_{j}^{\dagger}\right]=0.
\end{equation}
We can introduce the continuous magnon field $\hat{\psi}(\mathbf{r})$,
\begin{equation}
\hat{b}_{i}=\frac{1}{\sqrt{V_{i}}}\int_{V_{i}}\hat{\psi}(\mathbf{r})d^{3}r,
\end{equation}
where $V_{i}$ is the volume of the unit cell associated with the
spin at lattice site $i$. The replacement of the magnon operators
$\hat{b}_{i}$ by the field $\hat{\psi}(\mathbf{r})$ is valid for
long-wavelength magnons (i.e., in the continuum limit). The normalization
is chosen such that the magnon field fulfills
\begin{equation}
\left[\hat{\psi}(\mathbf{r}),\hat{\psi}^{\dagger}(\mathbf{r}')\right]=\delta(\mathbf{r}-\mathbf{r}').
\end{equation}
This allows us to define the canonically conjugate field
\begin{equation}
\hat{\pi}(\mathbf{r})=i\hbar\hat{\psi}^{\dagger}(\mathbf{r}),
\end{equation}
with the canonical commutation relation
\begin{equation}
\left[\hat{\psi}(\mathbf{r}),\hat{\pi}(\mathbf{r}')\right]=i\hbar\delta(\mathbf{r}-\mathbf{r}').
\end{equation}
The corresponding classical magnon field fulfills then the canonical
Poisson bracket relations at equal times,
\begin{equation}
\left\{ \psi(\mathbf{r},t),\pi(\mathbf{r}',t)\right\} =\delta(\mathbf{r}-\mathbf{r}').
\end{equation}

Because a general spin Hamiltonian $\mathcal{H}[\psi,\pi]$ is a functional
of the fields and their gradients but not of their time derivatives,
we obtain the classical Lagrangian \cite{Greiner}
\begin{equation}
\mathcal{L}=\int\pi\dot{\psi}\,d^{3}r-\mathcal{H}\left[\psi,\pi\right].
\end{equation}
It is now straight-forward to apply Noether's theorem to this Lagrangian
\cite{Greiner} and to derive the classical magnon pseudomomentum
and pseudoangular momentum,
\begin{align}
\tilde{\mathbf{P}}_{\text{m}} & =-i\hbar\int\psi^{*}\boldsymbol{\nabla}\psi\,d^{3}r,\label{eq:mP}\\
\tilde{\mathbf{L}}_{\text{m}} & =-i\hbar\int\mathbf{r}\times\left(\psi^{*}\boldsymbol{\nabla}\psi\right)d^{3}r.\label{eq:mL}
\end{align}
The magnon field $\psi(\mathbf{r})$ is a complex scalar field and
therefore the magnon pseudoangular momentum has only an orbital contribution
and no pseudospin. The pseudomomentum $\hbar\mathbf{k}$ of magnons
enters the conservation of pseudomomentum, which is for example important
in spin transfer processes \cite{Mitrofanov2020,Tramsen2021}. The
total spin angular momentum is determined by the number of excited
magnons,
\begin{align}
S_{\text{tot}}^{z} & =S_{\text{tot}}^{0}-\hbar\sum_{i}\langle\hat{n}_{i}\rangle\nonumber \\
 & =S_{\text{tot}}^{0}-\hbar\int\psi^{*}(\mathbf{r})\psi(\mathbf{r})d^{3}r,\label{eq:spin}
\end{align}
where $S_{\text{tot}}^{0}=\hbar NS$, with $N$ the number of lattice
sites, is the total spin if no magnons are excited. 

We note that it is also possible to derive the pseudoangular momentum
of the magnetization field $\mathbf{M}(\mathbf{r})$, which is a vector
field with a spin contribution to its pseudoangular momentum \cite{Tsukernik1984,Yan2013},
although there are complications with respect to the application of
Noether's theorem since the resulting pseudomomenta are gauge dependent
and not uniquely defined \cite{Papanicolaou1991,Yan2013,Tchernyshyov2015,Dasgupta2018}.
This is however not a fundamental issue because these quantities are
pseudomomenta and not proper momenta, which would have to be well-defined.
Similarly, the pseudomomentum and pseudoangular momentum of magnons
are only defined with respect to a given magnon quantization axis,
which is in principle arbitrary. However, magnons usually describe
fluctuations with respect to a given magnetic equilibrium configuration
that physically defines a preferred quantization axis for each spin.

By comparing our results for the magnon pseudomomentum and pseudoangular momentum, Eqs.~(\ref{eq:mP}-\ref{eq:mL}), with the magnon
momentum and magnon orbital angular momentum of Ref.~\cite{Jia2019},
we find that they are indeed the same quantities: the pseudomomentum
and pseudoangular momentum of magnons. Ref.~\cite{Jiang2019} uses
instead the definition of the pseudo-orbital angular momentum of the
magnetization field \cite{Tsukernik1984,Yan2013}.

Since a magnon excitation is only associated with a change of the
magnetization and therefore of the angular momentum, a magnon cannot
have any momentum, only pseudomomentum. The angular momentum associated
with the magnon field is determined by the number of magnons alone,
Eq.~(\ref{eq:spin}), and does not depend on the spatial structure
of the magnon field, whereas the pseudo-orbital angular momentum $\tilde{\mathbf{L}}_{\text{m}}$
shows such a dependence and does not correspond to any contribution
to the angular momentum. If the magnetization is not due to the electron
spin alone but does also have an orbital contribution, then there
is a correlation between the electron orbital angular momentum and
the number of magnons. 

We point out that the recently proposed orbital magnetic
moments of phonons \cite{Jurascheck2019} and magnons \cite{Neumann2020}
are not related to pseudoangular momentum. Magnetic moments are always
due to the angular momentum of charged particles. In the case of Ref.~\cite{Jurascheck2019}
the underlying angular momentum is the orbital angular momentum of
ions, and in the case of Ref.~\cite{Neumann2020} the orbital angular
momentum of electrons.

In summary, we have shown that the conservation of pseudoangular
momentum arises when a condensed matter system is invariant under
rotations of the emergent fields that are associated with quasiparticles.
We have considered the pseudoangular momentum of phonons and magnons,
which are derived from Noether's theorem, and demonstrated that the
recently discussed orbital angular momenta of phonons \cite{Ayub2011,Nakane2018,Mentink2019}
and magnons \cite{Jia2019,Jiang2019} are in fact pseudoangular momenta.
In general, we conclude that when Noether's theorem is applied to
emergent quasiparticle fields, the resulting conserved momenta are
pseudomomenta. The examples discussed here show that it is important
to consider the difference between angular momentum and pseudoangular
momentum, which is especially important in spintronics applications
where angular momentum is used as an information carrier. While angular
momentum is in principle always conserved in a closed system, pseudoangular momentum is not exactly conserved in real materials, which
should be kept in mind for potential experimental applications. The
distinction between angular momentum and pseudoangular momentum could
also be crucial for the analysis of the angular momentum associated
with the pseudospin in graphene, which has been argued to represent
a real angular momentum \cite{Mecklenburg2011,Song2015}.
\\

I thank Charles Sebens for discussions on angular momentum
and spin, Mikhael Katsnelson and Johan Mentink for discussing the
concept of pseudomomentum, and Jorge E. Hirsch for discussions on
the transfer of angular momentum in superconductors and ferromagnets
and his hospitality after the cancellation of the APS March Meeting
2020. Special thanks go to Anna Delin, Danny Thonig, and Olle Eriksson
for feedback on this research project. Financial support from the
Knut and Alice Wallenberg Foundation through Grant No. 2018.0060 is
gratefully acknowledged.


\begin{thebibliography}{64}%
\makeatletter
\providecommand \@ifxundefined [1]{%
 \@ifx{#1\undefined}
}%
\providecommand \@ifnum [1]{%
 \ifnum #1\expandafter \@firstoftwo
 \else \expandafter \@secondoftwo
 \fi
}%
\providecommand \@ifx [1]{%
 \ifx #1\expandafter \@firstoftwo
 \else \expandafter \@secondoftwo
 \fi
}%
\providecommand \natexlab [1]{#1}%
\providecommand \enquote  [1]{``#1''}%
\providecommand \bibnamefont  [1]{#1}%
\providecommand \bibfnamefont [1]{#1}%
\providecommand \citenamefont [1]{#1}%
\providecommand \href@noop [0]{\@secondoftwo}%
\providecommand \href [0]{\begingroup \@sanitize@url \@href}%
\providecommand \@href[1]{\@@startlink{#1}\@@href}%
\providecommand \@@href[1]{\endgroup#1\@@endlink}%
\providecommand \@sanitize@url [0]{\catcode `\\12\catcode `\$12\catcode
  `\&12\catcode `\#12\catcode `\^12\catcode `\_12\catcode `\%12\relax}%
\providecommand \@@startlink[1]{}%
\providecommand \@@endlink[0]{}%
\providecommand \url  [0]{\begingroup\@sanitize@url \@url }%
\providecommand \@url [1]{\endgroup\@href {#1}{\urlprefix }}%
\providecommand \urlprefix  [0]{URL }%
\providecommand \Eprint [0]{\href }%
\providecommand \doibase [0]{https://doi.org/}%
\providecommand \selectlanguage [0]{\@gobble}%
\providecommand \bibinfo  [0]{\@secondoftwo}%
\providecommand \bibfield  [0]{\@secondoftwo}%
\providecommand \translation [1]{[#1]}%
\providecommand \BibitemOpen [0]{}%
\providecommand \bibitemStop [0]{}%
\providecommand \bibitemNoStop [0]{.\EOS\space}%
\providecommand \EOS [0]{\spacefactor3000\relax}%
\providecommand \BibitemShut  [1]{\csname bibitem#1\endcsname}%
\let\auto@bib@innerbib\@empty
%</preamble>
\bibitem [{\citenamefont {Einstein}\ and\ \citenamefont
  {de~Haas}(1915)}]{Einstein15}%
  \BibitemOpen
  \bibfield  {author} {\bibinfo {author} {\bibfnamefont {A.}~\bibnamefont
  {Einstein}}\ and\ \bibinfo {author} {\bibfnamefont {W.~J.}\ \bibnamefont
  {de~Haas}},\ }\bibfield  {title} {\bibinfo {title} {{Experimental proof of
  the existence of Amp\`ere's molecular currents}},\ }\href
  {http://www.dwc.knaw.nl/DL/publications/PU00012546.pdf} {\bibfield  {journal}
  {\bibinfo  {journal} {Proc. KNAW}\ }\textbf {\bibinfo {volume} {18}},\ \bibinfo {pages} {696} (\bibinfo {year} {1915})}\BibitemShut {NoStop}%
\bibitem [{\citenamefont {Barnett}(1915)}]{Barnett1915}%
  \BibitemOpen
  \bibfield  {author} {\bibinfo {author} {\bibfnamefont {S.~J.}\ \bibnamefont
  {Barnett}},\ }\bibfield  {title} {\bibinfo {title} {Magnetization by
  rotation},\ }\href {https://doi.org/10.1103/PhysRev.6.239} {\bibfield
  {journal} {\bibinfo  {journal} {Phys. Rev.}\ }\textbf {\bibinfo {volume}
  {6}},\ \bibinfo {pages} {239} (\bibinfo {year} {1915})}\BibitemShut {NoStop}%
\bibitem [{\citenamefont {Noether}(1918)}]{Noether1918}%
  \BibitemOpen
  \bibfield  {author} {\bibinfo {author} {\bibfnamefont {E.}~\bibnamefont
  {Noether}},\ }\bibfield  {title} {\bibinfo {title} {{Invariante
  Variationsprobleme}},\ }\href@noop {} {\bibfield  {journal} {\bibinfo
  {journal} {Nachr. Ges. Wiss. G\"ottingen, Math.-Phys. Kl.}\ }\textbf
  {\bibinfo {volume} {2}},\ \bibinfo {pages} {235} (\bibinfo {year}
  {1918})}\BibitemShut {NoStop}%
\bibitem [{\citenamefont {Gordon}(1973)}]{Gordon1973}%
  \BibitemOpen
  \bibfield  {author} {\bibinfo {author} {\bibfnamefont {J.~P.}\ \bibnamefont
  {Gordon}},\ }\bibfield  {title} {\bibinfo {title} {Radiation forces and
  momenta in dielectric media},\ }\href {https://doi.org/10.1103/PhysRevA.8.14}
  {\bibfield  {journal} {\bibinfo  {journal} {Phys. Rev. A}\ }\textbf {\bibinfo
  {volume} {8}},\ \bibinfo {pages} {14} (\bibinfo {year} {1973})}\BibitemShut
  {NoStop}%
\bibitem [{\citenamefont {Peierls}(1976)}]{Peierls1976}%
  \BibitemOpen
  \bibfield  {author} {\bibinfo {author} {\bibfnamefont {R.}~\bibnamefont
  {Peierls}},\ }\bibfield  {title} {\bibinfo {title} {{The Momentum of Light in
  a Refracting Medium}},\ }\href {http://www.jstor.org/stable/79058} {\bibfield
   {journal} {\bibinfo  {journal} {Proc. R. Soc. A}\ }\textbf {\bibinfo
  {volume} {347}},\ \bibinfo {pages} {475} (\bibinfo {year}
  {1976})}\BibitemShut {NoStop}%
\bibitem [{\citenamefont {Peierls}(1979)}]{Peierls1979}%
  \BibitemOpen
  \bibfield  {author} {\bibinfo {author} {\bibfnamefont {R.}~\bibnamefont
  {Peierls}},\ }\href@noop {} {\emph {\bibinfo {title} {{Surprises in
  Theoretical Physics}}}}\ (\bibinfo  {publisher} {Princeton University
  Press},\ \bibinfo {year} {1979})\BibitemShut {NoStop}%
\bibitem [{\citenamefont {Thellung}(1980)}]{Thellung1980}%
  \BibitemOpen
  \bibfield  {author} {\bibinfo {author} {\bibfnamefont {A.}~\bibnamefont
  {Thellung}},\ }\bibfield  {title} {\bibinfo {title} {Two-fluid equations for
  phonons without momentum},\ }\href
  {https://doi.org/https://doi.org/10.1016/0003-4916(80)90100-1} {\bibfield
  {journal} {\bibinfo  {journal} {Ann. Phys. (N. Y.)}\ }\textbf {\bibinfo
  {volume} {127}},\ \bibinfo {pages} {289 } (\bibinfo {year}
  {1980})}\BibitemShut {NoStop}%
\bibitem [{\citenamefont {Mcintyre}(1981)}]{Mcintyre1981}%
  \BibitemOpen
  \bibfield  {author} {\bibinfo {author} {\bibfnamefont {M.~E.}\ \bibnamefont
  {Mcintyre}},\ }\bibfield  {title} {\bibinfo {title} {On the `wave momentum'
  myth},\ }\href {https://doi.org/10.1017/S0022112081001626} {\bibfield
  {journal} {\bibinfo  {journal} {J. Fluid Mech.}\ }\textbf {\bibinfo {volume}
  {106}},\ \bibinfo {pages} {331} (\bibinfo {year} {1981})}\BibitemShut
  {NoStop}%
\bibitem [{\citenamefont {Nelson}(1991)}]{Nelson1991}%
  \BibitemOpen
  \bibfield  {author} {\bibinfo {author} {\bibfnamefont {D.~F.}\ \bibnamefont
  {Nelson}},\ }\bibfield  {title} {\bibinfo {title} {{Momentum, pseudomomentum,
  and wave momentum: Toward resolving the Minkowski-Abraham controversy}},\
  }\href {https://doi.org/10.1103/PhysRevA.44.3985} {\bibfield  {journal}
  {\bibinfo  {journal} {Phys. Rev. A}\ }\textbf {\bibinfo {volume} {44}},\
  \bibinfo {pages} {3985} (\bibinfo {year} {1991})}\BibitemShut {NoStop}%
\bibitem [{\citenamefont {Peierls}(1991)}]{Peierls1991}%
  \BibitemOpen
  \bibfield  {author} {\bibinfo {author} {\bibfnamefont {R.}~\bibnamefont
  {Peierls}},\ }\href@noop {} {\emph {\bibinfo {title} {{More Surprises in
  Theoretical Physics}}}}\ (\bibinfo  {publisher} {Princeton University
  Press},\ \bibinfo {year} {1991})\BibitemShut {NoStop}%
\bibitem [{\citenamefont {Thellung}(1994)}]{Thellung1994}%
  \BibitemOpen
  \bibfield  {author} {\bibinfo {author} {\bibfnamefont {A.}~\bibnamefont
  {Thellung}},\ }\bibinfo {title} {{Momentum and Quasimomentum in the Physics
  of Condensed Matter}},\ in\ \href
  {https://doi.org/10.1007/978-1-4615-2455-7_2} {\emph {\bibinfo {booktitle}
  {{Die Kunst of Phonons: Lectures from the Winter School of Theoretical
  Physics}}}}\ (\bibinfo  {publisher} {Springer US},\ \bibinfo {address}
  {Boston, MA},\ \bibinfo {year} {1994})\BibitemShut {NoStop}%
\bibitem [{\citenamefont {Stone}(2002)}]{Stone2002}%
  \BibitemOpen
  \bibfield  {author} {\bibinfo {author} {\bibfnamefont {M.}~\bibnamefont
  {Stone}},\ }\bibinfo {title} {{Phonons and forces: Momentum {\it versus}
  pseudomomentum in moving fluids}},\ in\ \href
  {https://doi.org/10.1142/9789812778178_0013} {\emph {\bibinfo {booktitle}
  {{Artificial Black Holes}}}}\ (\bibinfo  {publisher} {World Scientific},\
  \bibinfo {address} {Singapore},\ \bibinfo {year} {2002})\BibitemShut
  {NoStop}%
\bibitem [{\citenamefont {Heller}\ and\ \citenamefont
  {Kim}(2019)}]{Heller2019}%
  \BibitemOpen
  \bibfield  {author} {\bibinfo {author} {\bibfnamefont {E.~J.}\ \bibnamefont
  {Heller}}\ and\ \bibinfo {author} {\bibfnamefont {D.}~\bibnamefont {Kim}},\
  }\bibfield  {title} {\bibinfo {title} {{Schr\"odinger Correspondence Applied
  to Crystals}},\ }\href {https://doi.org/10.1021/acs.jpca.8b11746} {\bibfield
  {journal} {\bibinfo  {journal} {J. Phys. Chem. A}\ }\textbf {\bibinfo
  {volume} {123}},\ \bibinfo {pages} {4379} (\bibinfo {year}
  {2019})}\BibitemShut {NoStop}%
\bibitem [{\citenamefont {Pfeifer}\ \emph {et~al.}(2007)\citenamefont
  {Pfeifer}, \citenamefont {Nieminen}, \citenamefont {Heckenberg},\ and\
  \citenamefont {Rubinsztein-Dunlop}}]{Pfeifer2007}%
  \BibitemOpen
  \bibfield  {author} {\bibinfo {author} {\bibfnamefont {R.~N.~C.}\
  \bibnamefont {Pfeifer}}, \bibinfo {author} {\bibfnamefont {T.~A.}\
  \bibnamefont {Nieminen}}, \bibinfo {author} {\bibfnamefont {N.~R.}\
  \bibnamefont {Heckenberg}},\ and\ \bibinfo {author} {\bibfnamefont
  {H.}~\bibnamefont {Rubinsztein-Dunlop}},\ }\bibfield  {title} {\bibinfo
  {title} {Colloquium: Momentum of an electromagnetic wave in dielectric
  media},\ }\href {https://doi.org/10.1103/RevModPhys.79.1197} {\bibfield
  {journal} {\bibinfo  {journal} {Rev. Mod. Phys.}\ }\textbf {\bibinfo {volume}
  {79}},\ \bibinfo {pages} {1197} (\bibinfo {year} {2007})}\BibitemShut
  {NoStop}%
\bibitem [{\citenamefont {Thomas}\ and\ \citenamefont
  {Marchiano}(2003)}]{Thomas2003}%
  \BibitemOpen
  \bibfield  {author} {\bibinfo {author} {\bibfnamefont {J.-L.}\ \bibnamefont
  {Thomas}}\ and\ \bibinfo {author} {\bibfnamefont {R.}~\bibnamefont
  {Marchiano}},\ }\bibfield  {title} {\bibinfo {title} {{Pseudo Angular
  Momentum and Topological Charge Conservation for Nonlinear Acoustical
  Vortices}},\ }\href {https://doi.org/10.1103/PhysRevLett.91.244302}
  {\bibfield  {journal} {\bibinfo  {journal} {Phys. Rev. Lett.}\ }\textbf
  {\bibinfo {volume} {91}},\ \bibinfo {pages} {244302} (\bibinfo {year}
  {2003})}\BibitemShut {NoStop}%
\bibitem [{\citenamefont {Zhang}\ and\ \citenamefont {Niu}(2015)}]{Zhang2015}%
  \BibitemOpen
  \bibfield  {author} {\bibinfo {author} {\bibfnamefont {L.}~\bibnamefont
  {Zhang}}\ and\ \bibinfo {author} {\bibfnamefont {Q.}~\bibnamefont {Niu}},\
  }\bibfield  {title} {\bibinfo {title} {{Chiral Phonons at High-Symmetry
  Points in Monolayer Hexagonal Lattices}},\ }\href
  {https://doi.org/10.1103/PhysRevLett.115.115502} {\bibfield  {journal}
  {\bibinfo  {journal} {Phys. Rev. Lett.}\ }\textbf {\bibinfo {volume} {115}},\
  \bibinfo {pages} {115502} (\bibinfo {year} {2015})}\BibitemShut {NoStop}%
\bibitem [{\citenamefont {Zhu}\ \emph {et~al.}(2018)\citenamefont {Zhu},
  \citenamefont {Yi}, \citenamefont {Li}, \citenamefont {Xiao}, \citenamefont
  {Zhang}, \citenamefont {Yang}, \citenamefont {Kaindl}, \citenamefont {Li},
  \citenamefont {Wang},\ and\ \citenamefont {Zhang}}]{Zhu2018}%
  \BibitemOpen
  \bibfield  {author} {\bibinfo {author} {\bibfnamefont {H.}~\bibnamefont
  {Zhu}}, \bibinfo {author} {\bibfnamefont {J.}~\bibnamefont {Yi}}, \bibinfo
  {author} {\bibfnamefont {M.-Y.}\ \bibnamefont {Li}}, \bibinfo {author}
  {\bibfnamefont {J.}~\bibnamefont {Xiao}}, \bibinfo {author} {\bibfnamefont
  {L.}~\bibnamefont {Zhang}}, \bibinfo {author} {\bibfnamefont {C.-W.}\
  \bibnamefont {Yang}}, \bibinfo {author} {\bibfnamefont {R.~A.}\ \bibnamefont
  {Kaindl}}, \bibinfo {author} {\bibfnamefont {L.-J.}\ \bibnamefont {Li}},
  \bibinfo {author} {\bibfnamefont {Y.}~\bibnamefont {Wang}},\ and\ \bibinfo
  {author} {\bibfnamefont {X.}~\bibnamefont {Zhang}},\ }\bibfield  {title}
  {\bibinfo {title} {Observation of chiral phonons},\ }\href
  {https://doi.org/10.1126/science.aar2711} {\bibfield  {journal} {\bibinfo
  {journal} {Science}\ }\textbf {\bibinfo {volume} {359}},\ \bibinfo {pages}
  {579} (\bibinfo {year} {2018})}\BibitemShut {NoStop}%
\bibitem [{\citenamefont {Tatsumi}\ \emph {et~al.}(2018)\citenamefont
  {Tatsumi}, \citenamefont {Kaneko},\ and\ \citenamefont
  {Saito}}]{Tatsumi2018}%
  \BibitemOpen
  \bibfield  {author} {\bibinfo {author} {\bibfnamefont {Y.}~\bibnamefont
  {Tatsumi}}, \bibinfo {author} {\bibfnamefont {T.}~\bibnamefont {Kaneko}},\
  and\ \bibinfo {author} {\bibfnamefont {R.}~\bibnamefont {Saito}},\ }\bibfield
   {title} {\bibinfo {title} {{Conservation law of angular momentum in
  helicity-dependent Raman and Rayleigh scattering}},\ }\href
  {https://doi.org/10.1103/PhysRevB.97.195444} {\bibfield  {journal} {\bibinfo
  {journal} {Phys. Rev. B}\ }\textbf {\bibinfo {volume} {97}},\ \bibinfo
  {pages} {195444} (\bibinfo {year} {2018})}\BibitemShut {NoStop}%
\bibitem [{\citenamefont {Zhang}\ \emph
  {et~al.}(2020{\natexlab{a}})\citenamefont {Zhang}, \citenamefont
  {Srivastava}, \citenamefont {Li},\ and\ \citenamefont {Zhang}}]{ZhangW2020}%
  \BibitemOpen
  \bibfield  {author} {\bibinfo {author} {\bibfnamefont {W.}~\bibnamefont
  {Zhang}}, \bibinfo {author} {\bibfnamefont {A.}~\bibnamefont {Srivastava}},
  \bibinfo {author} {\bibfnamefont {X.}~\bibnamefont {Li}},\ and\ \bibinfo
  {author} {\bibfnamefont {L.}~\bibnamefont {Zhang}},\ }\bibfield  {title}
  {\bibinfo {title} {{Chiral phonons in the indirect optical transition of a
  ${\mathrm{MoS}}_{2}/{\mathrm{WS}}_{2}$ heterostructure}},\ }\href
  {https://doi.org/10.1103/PhysRevB.102.174301} {\bibfield  {journal} {\bibinfo
   {journal} {Phys. Rev. B}\ }\textbf {\bibinfo {volume} {102}},\ \bibinfo
  {pages} {174301} (\bibinfo {year} {2020}{\natexlab{a}})}\BibitemShut
  {NoStop}%
\bibitem [{\citenamefont {Garanin}\ and\ \citenamefont
  {Chudnovsky}(2015)}]{Garanin2015}%
  \BibitemOpen
  \bibfield  {author} {\bibinfo {author} {\bibfnamefont {D.~A.}\ \bibnamefont
  {Garanin}}\ and\ \bibinfo {author} {\bibfnamefont {E.~M.}\ \bibnamefont
  {Chudnovsky}},\ }\bibfield  {title} {\bibinfo {title} {Angular momentum in
  spin-phonon processes},\ }\href {https://doi.org/10.1103/PhysRevB.92.024421}
  {\bibfield  {journal} {\bibinfo  {journal} {Phys. Rev. B}\ }\textbf {\bibinfo
  {volume} {92}},\ \bibinfo {pages} {024421} (\bibinfo {year}
  {2015})}\BibitemShut {NoStop}%
\bibitem [{\citenamefont {Nakane}\ and\ \citenamefont
  {Kohno}(2018)}]{Nakane2018}%
  \BibitemOpen
  \bibfield  {author} {\bibinfo {author} {\bibfnamefont {J.~J.}\ \bibnamefont
  {Nakane}}\ and\ \bibinfo {author} {\bibfnamefont {H.}~\bibnamefont {Kohno}},\
  }\bibfield  {title} {\bibinfo {title} {Angular momentum of phonons and its
  application to single-spin relaxation},\ }\href
  {https://doi.org/10.1103/PhysRevB.97.174403} {\bibfield  {journal} {\bibinfo
  {journal} {Phys. Rev. B}\ }\textbf {\bibinfo {volume} {97}},\ \bibinfo
  {pages} {174403} (\bibinfo {year} {2018})}\BibitemShut {NoStop}%
\bibitem [{\citenamefont {Mentink}\ \emph {et~al.}(2019)\citenamefont
  {Mentink}, \citenamefont {Katsnelson},\ and\ \citenamefont
  {Lemeshko}}]{Mentink2019}%
  \BibitemOpen
  \bibfield  {author} {\bibinfo {author} {\bibfnamefont {J.~H.}\ \bibnamefont
  {Mentink}}, \bibinfo {author} {\bibfnamefont {M.~I.}\ \bibnamefont
  {Katsnelson}},\ and\ \bibinfo {author} {\bibfnamefont {M.}~\bibnamefont
  {Lemeshko}},\ }\bibfield  {title} {\bibinfo {title} {{Quantum many-body
  dynamics of the Einstein--de Haas effect}},\ }\href
  {https://doi.org/10.1103/PhysRevB.99.064428} {\bibfield  {journal} {\bibinfo
  {journal} {Phys. Rev. B}\ }\textbf {\bibinfo {volume} {99}},\ \bibinfo
  {pages} {064428} (\bibinfo {year} {2019})}\BibitemShut {NoStop}%
\bibitem [{\citenamefont {Streib}\ \emph {et~al.}(2018)\citenamefont {Streib},
  \citenamefont {Keshtgar},\ and\ \citenamefont {Bauer}}]{Streib2018}%
  \BibitemOpen
  \bibfield  {author} {\bibinfo {author} {\bibfnamefont {S.}~\bibnamefont
  {Streib}}, \bibinfo {author} {\bibfnamefont {H.}~\bibnamefont {Keshtgar}},\
  and\ \bibinfo {author} {\bibfnamefont {G.~E.~W.}\ \bibnamefont {Bauer}},\
  }\bibfield  {title} {\bibinfo {title} {Damping of magnetization dynamics by
  phonon pumping},\ }\href {https://doi.org/10.1103/PhysRevLett.121.027202}
  {\bibfield  {journal} {\bibinfo  {journal} {Phys. Rev. Lett.}\ }\textbf
  {\bibinfo {volume} {121}},\ \bibinfo {pages} {027202} (\bibinfo {year}
  {2018})}\BibitemShut {NoStop}%
\bibitem [{\citenamefont {R\"uckriegel}\ \emph {et~al.}(2020)\citenamefont
  {R\"uckriegel}, \citenamefont {Streib}, \citenamefont {Bauer},\ and\
  \citenamefont {Duine}}]{Rueckiegel2020}%
  \BibitemOpen
  \bibfield  {author} {\bibinfo {author} {\bibfnamefont {A.}~\bibnamefont
  {R\"uckriegel}}, \bibinfo {author} {\bibfnamefont {S.}~\bibnamefont
  {Streib}}, \bibinfo {author} {\bibfnamefont {G.~E.~W.}\ \bibnamefont
  {Bauer}},\ and\ \bibinfo {author} {\bibfnamefont {R.~A.}\ \bibnamefont
  {Duine}},\ }\bibfield  {title} {\bibinfo {title} {Angular momentum
  conservation and phonon spin in magnetic insulators},\ }\href
  {https://doi.org/10.1103/PhysRevB.101.104402} {\bibfield  {journal} {\bibinfo
   {journal} {Phys. Rev. B}\ }\textbf {\bibinfo {volume} {101}},\ \bibinfo
  {pages} {104402} (\bibinfo {year} {2020})}\BibitemShut {NoStop}%
\bibitem [{\citenamefont {R\"uckriegel}\ and\ \citenamefont
  {Duine}(2020)}]{Rueckriegel2020b}%
  \BibitemOpen
  \bibfield  {author} {\bibinfo {author} {\bibfnamefont {A.}~\bibnamefont
  {R\"uckriegel}}\ and\ \bibinfo {author} {\bibfnamefont {R.~A.}\ \bibnamefont
  {Duine}},\ }\bibfield  {title} {\bibinfo {title} {Long-range phonon spin
  transport in ferromagnet--nonmagnetic insulator heterostructures},\ }\href
  {https://doi.org/10.1103/PhysRevLett.124.117201} {\bibfield  {journal}
  {\bibinfo  {journal} {Phys. Rev. Lett.}\ }\textbf {\bibinfo {volume} {124}},\
  \bibinfo {pages} {117201} (\bibinfo {year} {2020})}\BibitemShut {NoStop}%
\bibitem [{\citenamefont {Zhang}\ \emph
  {et~al.}(2020{\natexlab{b}})\citenamefont {Zhang}, \citenamefont {Bauer},\
  and\ \citenamefont {Yu}}]{Zhang2020}%
  \BibitemOpen
  \bibfield  {author} {\bibinfo {author} {\bibfnamefont {X.}~\bibnamefont
  {Zhang}}, \bibinfo {author} {\bibfnamefont {G.~E.~W.}\ \bibnamefont
  {Bauer}},\ and\ \bibinfo {author} {\bibfnamefont {T.}~\bibnamefont {Yu}},\
  }\bibfield  {title} {\bibinfo {title} {Unidirectional pumping of phonons by
  magnetization dynamics},\ }\href
  {https://doi.org/10.1103/PhysRevLett.125.077203} {\bibfield  {journal}
  {\bibinfo  {journal} {Phys. Rev. Lett.}\ }\textbf {\bibinfo {volume} {125}},\
  \bibinfo {pages} {077203} (\bibinfo {year} {2020}{\natexlab{b}})}\BibitemShut
  {NoStop}%
\bibitem [{\citenamefont {Juraschek}\ \emph {et~al.}(2020)\citenamefont
  {Juraschek}, \citenamefont {Narang},\ and\ \citenamefont
  {Spaldin}}]{Juraschek2020}%
  \BibitemOpen
  \bibfield  {author} {\bibinfo {author} {\bibfnamefont {D.~M.}\ \bibnamefont
  {Juraschek}}, \bibinfo {author} {\bibfnamefont {P.}~\bibnamefont {Narang}},\
  and\ \bibinfo {author} {\bibfnamefont {N.~A.}\ \bibnamefont {Spaldin}},\
  }\bibfield  {title} {\bibinfo {title} {Phono-magnetic analogs to
  opto-magnetic effects},\ }\href
  {https://doi.org/10.1103/PhysRevResearch.2.043035} {\bibfield  {journal}
  {\bibinfo  {journal} {Phys. Rev. Research}\ }\textbf {\bibinfo {volume}
  {2}},\ \bibinfo {pages} {043035} (\bibinfo {year} {2020})}\BibitemShut
  {NoStop}%
\bibitem [{\citenamefont {Vonsovskii}\ and\ \citenamefont
  {Svirskii}(1962)}]{Vonsovskii1962}%
  \BibitemOpen
  \bibfield  {author} {\bibinfo {author} {\bibfnamefont {S.~V.}\ \bibnamefont
  {Vonsovskii}}\ and\ \bibinfo {author} {\bibfnamefont {M.~S.}\ \bibnamefont
  {Svirskii}},\ }\bibfield  {title} {\bibinfo {title} {{Phonon Spin}},\
  }\href@noop {} {\bibfield  {journal} {\bibinfo  {journal} {Sov. Phys. Solid
  State}\ }\textbf {\bibinfo {volume} {3}},\ \bibinfo {pages} {1568} (\bibinfo
  {year} {1962})}\BibitemShut {NoStop}%
\bibitem [{\citenamefont {Zhang}\ and\ \citenamefont {Niu}(2014)}]{Zhang2014}%
  \BibitemOpen
  \bibfield  {author} {\bibinfo {author} {\bibfnamefont {L.}~\bibnamefont
  {Zhang}}\ and\ \bibinfo {author} {\bibfnamefont {Q.}~\bibnamefont {Niu}},\
  }\bibfield  {title} {\bibinfo {title} {{Angular Momentum of Phonons and the
  Einstein--de Haas Effect}},\ }\href
  {https://doi.org/10.1103/PhysRevLett.112.085503} {\bibfield  {journal}
  {\bibinfo  {journal} {Phys. Rev. Lett.}\ }\textbf {\bibinfo {volume} {112}},\
  \bibinfo {pages} {085503} (\bibinfo {year} {2014})}\BibitemShut {NoStop}%
\bibitem [{\citenamefont {Holanda}\ \emph {et~al.}(2018)\citenamefont
  {Holanda}, \citenamefont {Maior}, \citenamefont {Azevedo},\ and\
  \citenamefont {Rezende}}]{Holanda2018}%
  \BibitemOpen
  \bibfield  {author} {\bibinfo {author} {\bibfnamefont {J.}~\bibnamefont
  {Holanda}}, \bibinfo {author} {\bibfnamefont {D.~S.}\ \bibnamefont {Maior}},
  \bibinfo {author} {\bibfnamefont {A.}~\bibnamefont {Azevedo}},\ and\ \bibinfo
  {author} {\bibfnamefont {S.~M.}\ \bibnamefont {Rezende}},\ }\bibfield
  {title} {\bibinfo {title} {Detecting the phonon spin in magnon-phonon
  conversion experiments},\ }\href {https://doi.org/10.1038/s41567-018-0079-y}
  {\bibfield  {journal} {\bibinfo  {journal} {Nat. Phys.}\ }\textbf {\bibinfo
  {volume} {14}},\ \bibinfo {pages} {500} (\bibinfo {year} {2018})}\BibitemShut
  {NoStop}%
\bibitem [{\citenamefont {An}\ \emph {et~al.}(2020)\citenamefont {An},
  \citenamefont {Litvinenko}, \citenamefont {Kohno}, \citenamefont {Fuad},
  \citenamefont {Naletov}, \citenamefont {Vila}, \citenamefont {Ebels},
  \citenamefont {de~Loubens}, \citenamefont {Hurdequint}, \citenamefont
  {Beaulieu}, \citenamefont {Ben~Youssef}, \citenamefont {Vukadinovic},
  \citenamefont {Bauer}, \citenamefont {Slavin}, \citenamefont {Tiberkevich},\
  and\ \citenamefont {Klein}}]{An2020}%
  \BibitemOpen
  \bibfield  {author} {\bibinfo {author} {\bibfnamefont {K.}~\bibnamefont
  {An}}, \bibinfo {author} {\bibfnamefont {A.~N.}\ \bibnamefont {Litvinenko}},
  \bibinfo {author} {\bibfnamefont {R.}~\bibnamefont {Kohno}}, \bibinfo
  {author} {\bibfnamefont {A.~A.}\ \bibnamefont {Fuad}}, \bibinfo {author}
  {\bibfnamefont {V.~V.}\ \bibnamefont {Naletov}}, \bibinfo {author}
  {\bibfnamefont {L.}~\bibnamefont {Vila}}, \bibinfo {author} {\bibfnamefont
  {U.}~\bibnamefont {Ebels}}, \bibinfo {author} {\bibfnamefont
  {G.}~\bibnamefont {de~Loubens}}, \bibinfo {author} {\bibfnamefont
  {H.}~\bibnamefont {Hurdequint}}, \bibinfo {author} {\bibfnamefont
  {N.}~\bibnamefont {Beaulieu}}, \bibinfo {author} {\bibfnamefont
  {J.}~\bibnamefont {Ben~Youssef}}, \bibinfo {author} {\bibfnamefont
  {N.}~\bibnamefont {Vukadinovic}}, \bibinfo {author} {\bibfnamefont
  {G.~E.~W.}\ \bibnamefont {Bauer}}, \bibinfo {author} {\bibfnamefont {A.~N.}\
  \bibnamefont {Slavin}}, \bibinfo {author} {\bibfnamefont {V.~S.}\
  \bibnamefont {Tiberkevich}},\ and\ \bibinfo {author} {\bibfnamefont
  {O.}~\bibnamefont {Klein}},\ }\bibfield  {title} {\bibinfo {title} {Coherent
  long-range transfer of angular momentum between magnon kittel modes by
  phonons},\ }\href {https://doi.org/10.1103/PhysRevB.101.060407} {\bibfield
  {journal} {\bibinfo  {journal} {Phys. Rev. B}\ }\textbf {\bibinfo {volume}
  {101}},\ \bibinfo {pages} {060407(R)} (\bibinfo {year} {2020})}\BibitemShut
  {NoStop}%
\bibitem [{\citenamefont {Brataas}\ \emph {et~al.}(2020)\citenamefont
  {Brataas}, \citenamefont {{van Wees}}, \citenamefont {Klein}, \citenamefont
  {{de Loubens}},\ and\ \citenamefont {Viret}}]{Brataas2020}%
  \BibitemOpen
  \bibfield  {author} {\bibinfo {author} {\bibfnamefont {A.}~\bibnamefont
  {Brataas}}, \bibinfo {author} {\bibfnamefont {B.}~\bibnamefont {{van Wees}}},
  \bibinfo {author} {\bibfnamefont {O.}~\bibnamefont {Klein}}, \bibinfo
  {author} {\bibfnamefont {G.}~\bibnamefont {{de Loubens}}},\ and\ \bibinfo
  {author} {\bibfnamefont {M.}~\bibnamefont {Viret}},\ }\bibfield  {title}
  {\bibinfo {title} {Spin insulatronics},\ }\href
  {https://doi.org/https://doi.org/10.1016/j.physrep.2020.08.006} {\bibfield
  {journal} {\bibinfo  {journal} {Phys. Rep.}\ }\textbf {\bibinfo {volume}
  {885}},\ \bibinfo {pages} {1 } (\bibinfo {year} {2020})}\BibitemShut
  {NoStop}%
\bibitem [{\citenamefont {Ayub}\ \emph {et~al.}(2011)\citenamefont {Ayub},
  \citenamefont {Ali},\ and\ \citenamefont {Mendonca}}]{Ayub2011}%
  \BibitemOpen
  \bibfield  {author} {\bibinfo {author} {\bibfnamefont {M.~K.}\ \bibnamefont
  {Ayub}}, \bibinfo {author} {\bibfnamefont {S.}~\bibnamefont {Ali}},\ and\
  \bibinfo {author} {\bibfnamefont {J.~T.}\ \bibnamefont {Mendonca}},\
  }\bibfield  {title} {\bibinfo {title} {Phonons with orbital angular
  momentum},\ }\href {https://doi.org/10.1063/1.3655429} {\bibfield  {journal}
  {\bibinfo  {journal} {Phys. Plasmas}\ }\textbf {\bibinfo {volume} {18}},\
  \bibinfo {pages} {102117} (\bibinfo {year} {2011})}\BibitemShut {NoStop}%
\bibitem [{\citenamefont {Jia}\ \emph {et~al.}(2019)\citenamefont {Jia},
  \citenamefont {Ma}, \citenamefont {Sch{\"a}ffer},\ and\ \citenamefont
  {Berakdar}}]{Jia2019}%
  \BibitemOpen
  \bibfield  {author} {\bibinfo {author} {\bibfnamefont {C.}~\bibnamefont
  {Jia}}, \bibinfo {author} {\bibfnamefont {D.}~\bibnamefont {Ma}}, \bibinfo
  {author} {\bibfnamefont {A.~F.}\ \bibnamefont {Sch{\"a}ffer}},\ and\ \bibinfo
  {author} {\bibfnamefont {J.}~\bibnamefont {Berakdar}},\ }\bibfield  {title}
  {\bibinfo {title} {Twisted magnon beams carrying orbital angular momentum},\
  }\href {https://doi.org/10.1038/s41467-019-10008-3} {\bibfield  {journal}
  {\bibinfo  {journal} {Nat. Commun.}\ }\textbf {\bibinfo {volume} {10}},\
  \bibinfo {pages} {2077} (\bibinfo {year} {2019})}\BibitemShut {NoStop}%
\bibitem [{\citenamefont {Jiang}\ \emph {et~al.}(2020)\citenamefont {Jiang},
  \citenamefont {Yuan}, \citenamefont {Li}, \citenamefont {Wang}, \citenamefont
  {Zhang}, \citenamefont {Cao},\ and\ \citenamefont {Yan}}]{Jiang2019}%
  \BibitemOpen
  \bibfield  {author} {\bibinfo {author} {\bibfnamefont {Y.}~\bibnamefont
  {Jiang}}, \bibinfo {author} {\bibfnamefont {H.~Y.}\ \bibnamefont {Yuan}},
  \bibinfo {author} {\bibfnamefont {Z.-X.}\ \bibnamefont {Li}}, \bibinfo
  {author} {\bibfnamefont {Z.}~\bibnamefont {Wang}}, \bibinfo {author}
  {\bibfnamefont {H.~W.}\ \bibnamefont {Zhang}}, \bibinfo {author}
  {\bibfnamefont {Y.}~\bibnamefont {Cao}},\ and\ \bibinfo {author}
  {\bibfnamefont {P.}~\bibnamefont {Yan}},\ }\bibfield  {title} {\bibinfo
  {title} {Twisted magnon as a magnetic tweezer},\ }\href
  {https://doi.org/10.1103/PhysRevLett.124.217204} {\bibfield  {journal}
  {\bibinfo  {journal} {Phys. Rev. Lett.}\ }\textbf {\bibinfo {volume} {124}},\
  \bibinfo {pages} {217204} (\bibinfo {year} {2020})}\BibitemShut {NoStop}%
\bibitem [{\citenamefont {Yang}\ \emph {et~al.}(2018)\citenamefont {Yang},
  \citenamefont {Yang}, \citenamefont {Cao},\ and\ \citenamefont
  {Yan}}]{Yang2018}%
  \BibitemOpen
  \bibfield  {author} {\bibinfo {author} {\bibfnamefont {W.}~\bibnamefont
  {Yang}}, \bibinfo {author} {\bibfnamefont {H.}~\bibnamefont {Yang}}, \bibinfo
  {author} {\bibfnamefont {Y.}~\bibnamefont {Cao}},\ and\ \bibinfo {author}
  {\bibfnamefont {P.}~\bibnamefont {Yan}},\ }\bibfield  {title} {\bibinfo
  {title} {Photonic orbital angular momentum transfer and magnetic skyrmion
  rotation},\ }\href {https://doi.org/10.1364/OE.26.008778} {\bibfield
  {journal} {\bibinfo  {journal} {Opt. Express}\ }\textbf {\bibinfo {volume}
  {26}},\ \bibinfo {pages} {8778} (\bibinfo {year} {2018})}\BibitemShut
  {NoStop}%
\bibitem [{\citenamefont {Ashcroft}\ and\ \citenamefont
  {Mermin}(1976)}]{Ashcroft}%
  \BibitemOpen
  \bibfield  {author} {\bibinfo {author} {\bibfnamefont {N.}~\bibnamefont
  {Ashcroft}}\ and\ \bibinfo {author} {\bibfnamefont {N.}~\bibnamefont
  {Mermin}},\ }\href@noop {} {\emph {\bibinfo {title} {{Solid State
  Physics}}}}\ (\bibinfo  {publisher} {Saunders College Publishing},\ \bibinfo
  {address} {Philadelphia},\ \bibinfo {year} {1976})\BibitemShut {NoStop}%
\bibitem [{\citenamefont {Kittel}(1996)}]{Kittel}%
  \BibitemOpen
  \bibfield  {author} {\bibinfo {author} {\bibfnamefont {C.}~\bibnamefont
  {Kittel}},\ }\href@noop {} {\emph {\bibinfo {title} {{Introduction to Solid
  State Physics}}}},\ \bibinfo {edition} {7th}\ ed.\ (\bibinfo  {publisher}
  {Wiley},\ \bibinfo {address} {New York},\ \bibinfo {year} {1996})\BibitemShut
  {NoStop}%
\bibitem [{\citenamefont {Soper}(1976)}]{Soper}%
  \BibitemOpen
  \bibfield  {author} {\bibinfo {author} {\bibfnamefont {D.~E.}\ \bibnamefont
  {Soper}},\ }\href@noop {} {\emph {\bibinfo {title} {{Classical Field
  Theory}}}}\ (\bibinfo  {publisher} {Wiley},\ \bibinfo {address} {New York},\
  \bibinfo {year} {1976})\BibitemShut {NoStop}%
\bibitem [{\citenamefont {Greiner}\ and\ \citenamefont
  {Reinhardt}(1996)}]{Greiner}%
  \BibitemOpen
  \bibfield  {author} {\bibinfo {author} {\bibfnamefont {W.}~\bibnamefont
  {Greiner}}\ and\ \bibinfo {author} {\bibfnamefont {J.}~\bibnamefont
  {Reinhardt}},\ }\href {https://doi.org/10.1007/978-3-642-61485-9} {\emph
  {\bibinfo {title} {{Field Quantization}}}}\ (\bibinfo  {publisher}
  {Springer},\ \bibinfo {address} {Berlin},\ \bibinfo {year}
  {1996})\BibitemShut {NoStop}%
\bibitem [{\citenamefont {Cherepanov}\ \emph {et~al.}(1993)\citenamefont
  {Cherepanov}, \citenamefont {Kolokolov},\ and\ \citenamefont
  {L'vov}}]{Cherepanov1993}%
  \BibitemOpen
  \bibfield  {author} {\bibinfo {author} {\bibfnamefont {V.}~\bibnamefont
  {Cherepanov}}, \bibinfo {author} {\bibfnamefont {I.}~\bibnamefont
  {Kolokolov}},\ and\ \bibinfo {author} {\bibfnamefont {V.}~\bibnamefont
  {L'vov}},\ }\bibfield  {title} {\bibinfo {title} {{The saga of YIG: Spectra,
  thermodynamics, interaction and relaxation of magnons in a complex magnet}},\
  }\href {https://doi.org/10.1016/0370-1573(93)90107-O} {\bibfield  {journal}
  {\bibinfo  {journal} {Phys. Rep.}\ }\textbf {\bibinfo {volume} {229}},\
  \bibinfo {pages} {81 } (\bibinfo {year} {1993})}\BibitemShut {NoStop}%
\bibitem [{\citenamefont {Gurevich}\ and\ \citenamefont
  {Melkov}(1996)}]{Gurevich1996}%
  \BibitemOpen
  \bibfield  {author} {\bibinfo {author} {\bibfnamefont {A.~G.}\ \bibnamefont
  {Gurevich}}\ and\ \bibinfo {author} {\bibfnamefont {G.~A.}\ \bibnamefont
  {Melkov}},\ }\href@noop {} {\emph {\bibinfo {title} {{Magnetization
  Oscillations and Waves}}}}\ (\bibinfo  {publisher} {CRC},\ \bibinfo {address}
  {Boca Raton, FL},\ \bibinfo {year} {1996})\BibitemShut {NoStop}%
\bibitem [{\citenamefont {Maehrlein}(2017)}]{Maehrlein2016}%
  \BibitemOpen
  \bibfield  {author} {\bibinfo {author} {\bibfnamefont {S.}~\bibnamefont
  {Maehrlein}},\ }\emph {\bibinfo {title} {{Nonlinear Terahertz Phononics: A
  Novel Route to Controlling Matter}}},\ \href
  {https://refubium.fu-berlin.de/handle/fub188/9314} {Ph.D. thesis},\ \bibinfo
  {school} {Freie Universit\"at Berlin} (\bibinfo {year} {2017})\BibitemShut
  {NoStop}%
\bibitem [{\citenamefont {Maehrlein}\ \emph {et~al.}(2018)\citenamefont
  {Maehrlein}, \citenamefont {Radu}, \citenamefont {Maldonado}, \citenamefont
  {Paarmann}, \citenamefont {Gensch}, \citenamefont {Kalashnikova},
  \citenamefont {Pisarev}, \citenamefont {Wolf}, \citenamefont {Oppeneer},
  \citenamefont {Barker},\ and\ \citenamefont {Kampfrath}}]{Maehrlein2017}%
  \BibitemOpen
  \bibfield  {author} {\bibinfo {author} {\bibfnamefont {S.~F.}\ \bibnamefont
  {Maehrlein}}, \bibinfo {author} {\bibfnamefont {I.}~\bibnamefont {Radu}},
  \bibinfo {author} {\bibfnamefont {P.}~\bibnamefont {Maldonado}}, \bibinfo
  {author} {\bibfnamefont {A.}~\bibnamefont {Paarmann}}, \bibinfo {author}
  {\bibfnamefont {M.}~\bibnamefont {Gensch}}, \bibinfo {author} {\bibfnamefont
  {A.~M.}\ \bibnamefont {Kalashnikova}}, \bibinfo {author} {\bibfnamefont
  {R.~V.}\ \bibnamefont {Pisarev}}, \bibinfo {author} {\bibfnamefont
  {M.}~\bibnamefont {Wolf}}, \bibinfo {author} {\bibfnamefont {P.~M.}\
  \bibnamefont {Oppeneer}}, \bibinfo {author} {\bibfnamefont {J.}~\bibnamefont
  {Barker}},\ and\ \bibinfo {author} {\bibfnamefont {T.}~\bibnamefont
  {Kampfrath}},\ }\bibfield  {title} {\bibinfo {title} {Dissecting spin-phonon
  equilibration in ferrimagnetic insulators by ultrafast lattice excitation},\
  }\href {https://doi.org/10.1126/sciadv.aar5164} {\bibfield  {journal}
  {\bibinfo  {journal} {Sci. Adv.}\ }\textbf {\bibinfo {volume} {4}},\ \bibinfo
  {pages} {eaar5164} (\bibinfo {year} {2018})}\BibitemShut {NoStop}%
\bibitem [{\citenamefont {Kruglyak}\ \emph {et~al.}(2010)\citenamefont
  {Kruglyak}, \citenamefont {Demokritov},\ and\ \citenamefont
  {Grundler}}]{Kruglyak2010}%
  \BibitemOpen
  \bibfield  {author} {\bibinfo {author} {\bibfnamefont {V.~V.}\ \bibnamefont
  {Kruglyak}}, \bibinfo {author} {\bibfnamefont {S.~O.}\ \bibnamefont
  {Demokritov}},\ and\ \bibinfo {author} {\bibfnamefont {D.}~\bibnamefont
  {Grundler}},\ }\bibfield  {title} {\bibinfo {title} {Magnonics},\ }\href
  {http://stacks.iop.org/0022-3727/43/i=26/a=264001} {\bibfield  {journal}
  {\bibinfo  {journal} {J. Phys. D: Appl. Phys.}\ }\textbf {\bibinfo {volume}
  {43}},\ \bibinfo {pages} {264001} (\bibinfo {year} {2010})}\BibitemShut
  {NoStop}%
\bibitem [{\citenamefont {Chumak}\ \emph {et~al.}(2015)\citenamefont {Chumak},
  \citenamefont {Vasyuchka}, \citenamefont {Serga},\ and\ \citenamefont
  {Hillebrands}}]{Chumak2015}%
  \BibitemOpen
  \bibfield  {author} {\bibinfo {author} {\bibfnamefont {A.~V.}\ \bibnamefont
  {Chumak}}, \bibinfo {author} {\bibfnamefont {V.~I.}\ \bibnamefont
  {Vasyuchka}}, \bibinfo {author} {\bibfnamefont {A.~A.}\ \bibnamefont
  {Serga}},\ and\ \bibinfo {author} {\bibfnamefont {B.}~\bibnamefont
  {Hillebrands}},\ }\bibfield  {title} {\bibinfo {title} {Magnon spintronics},\
  }\href {http://dx.doi.org/10.1038/nphys3347} {\bibfield  {journal} {\bibinfo
  {journal} {Nat. Phys.}\ }\textbf {\bibinfo {volume} {11}},\ \bibinfo {pages}
  {453} (\bibinfo {year} {2015})}\BibitemShut {NoStop}%
\bibitem [{\citenamefont {Nikitov}\ \emph {et~al.}(2015)\citenamefont
  {Nikitov}, \citenamefont {Kalyabin}, \citenamefont {Lisenkov}, \citenamefont
  {Slavin}, \citenamefont {Barabanenkov}, \citenamefont {Osokin}, \citenamefont
  {Sadovnikov}, \citenamefont {Beginin}, \citenamefont {Morozova},
  \citenamefont {Filimonov}, \citenamefont {Khivintsev}, \citenamefont
  {Vysotsky}, \citenamefont {Sakharov},\ and\ \citenamefont
  {Pavlov}}]{Nikitov2015}%
  \BibitemOpen
  \bibfield  {author} {\bibinfo {author} {\bibfnamefont {S.~A.}\ \bibnamefont
  {Nikitov}}, \bibinfo {author} {\bibfnamefont {D.~V.}\ \bibnamefont
  {Kalyabin}}, \bibinfo {author} {\bibfnamefont {I.~V.}\ \bibnamefont
  {Lisenkov}}, \bibinfo {author} {\bibfnamefont {A.}~\bibnamefont {Slavin}},
  \bibinfo {author} {\bibfnamefont {Y.~N.}\ \bibnamefont {Barabanenkov}},
  \bibinfo {author} {\bibfnamefont {S.~A.}\ \bibnamefont {Osokin}}, \bibinfo
  {author} {\bibfnamefont {A.~V.}\ \bibnamefont {Sadovnikov}}, \bibinfo
  {author} {\bibfnamefont {E.~N.}\ \bibnamefont {Beginin}}, \bibinfo {author}
  {\bibfnamefont {M.~A.}\ \bibnamefont {Morozova}}, \bibinfo {author}
  {\bibfnamefont {Y.~A.}\ \bibnamefont {Filimonov}}, \bibinfo {author}
  {\bibfnamefont {Y.~V.}\ \bibnamefont {Khivintsev}}, \bibinfo {author}
  {\bibfnamefont {S.~L.}\ \bibnamefont {Vysotsky}}, \bibinfo {author}
  {\bibfnamefont {V.~K.}\ \bibnamefont {Sakharov}},\ and\ \bibinfo {author}
  {\bibfnamefont {E.~S.}\ \bibnamefont {Pavlov}},\ }\bibfield  {title}
  {\bibinfo {title} {Magnonics: a new research area in spintronics and spin
  wave electronics},\ }\href {http://stacks.iop.org/1063-7869/58/i=10/a=1002}
  {\bibfield  {journal} {\bibinfo  {journal} {Phys. Usp.}\ }\textbf {\bibinfo
  {volume} {58}},\ \bibinfo {pages} {1002} (\bibinfo {year}
  {2015})}\BibitemShut {NoStop}%
\bibitem [{\citenamefont {Ohanian}(1986)}]{Ohanian1986}%
  \BibitemOpen
  \bibfield  {author} {\bibinfo {author} {\bibfnamefont {H.~C.}\ \bibnamefont
  {Ohanian}},\ }\bibfield  {title} {\bibinfo {title} {What is spin?},\ }\href
  {https://doi.org/10.1119/1.14580} {\bibfield  {journal} {\bibinfo  {journal}
  {Am. J. Phys.}\ }\textbf {\bibinfo {volume} {54}},\ \bibinfo {pages} {500}
  (\bibinfo {year} {1986})}\BibitemShut {NoStop}%
\bibitem [{\citenamefont {Sebens}(2019)}]{Sebens2019}%
  \BibitemOpen
  \bibfield  {author} {\bibinfo {author} {\bibfnamefont {C.~T.}\ \bibnamefont
  {Sebens}},\ }\bibfield  {title} {\bibinfo {title} {How electrons spin},\
  }\href {https://doi.org/https://doi.org/10.1016/j.shpsb.2019.04.007}
  {\bibfield  {journal} {\bibinfo  {journal} {Stud. Hist. Philos. Sci. B}\
  }\textbf {\bibinfo {volume} {68}},\ \bibinfo {pages} {40 } (\bibinfo {year}
  {2019})}\BibitemShut {NoStop}%
\bibitem [{\citenamefont {Bliokh}\ \emph {et~al.}(2020)\citenamefont {Bliokh},
  \citenamefont {Punzmann}, \citenamefont {Xia}, \citenamefont {Nori},\ and\
  \citenamefont {Shats}}]{Bliokh2020}%
  \BibitemOpen
  \bibfield  {author} {\bibinfo {author} {\bibfnamefont {K.~Y.}\ \bibnamefont
  {Bliokh}}, \bibinfo {author} {\bibfnamefont {H.}~\bibnamefont {Punzmann}},
  \bibinfo {author} {\bibfnamefont {H.}~\bibnamefont {Xia}}, \bibinfo {author}
  {\bibfnamefont {F.}~\bibnamefont {Nori}},\ and\ \bibinfo {author}
  {\bibfnamefont {M.}~\bibnamefont {Shats}},\ }\href@noop {} {\bibinfo {title}
  {Relativistic field-theory spin and momentum in water waves}} (\bibinfo
  {year} {2020}),\ \Eprint {https://arxiv.org/abs/arXiv:2009.03245}
  {arXiv:2009.03245} \BibitemShut {NoStop}%
\bibitem [{\citenamefont {Lenstra}\ and\ \citenamefont
  {Mandel}(1982)}]{Lenstra1982}%
  \BibitemOpen
  \bibfield  {author} {\bibinfo {author} {\bibfnamefont {D.}~\bibnamefont
  {Lenstra}}\ and\ \bibinfo {author} {\bibfnamefont {L.}~\bibnamefont
  {Mandel}},\ }\bibfield  {title} {\bibinfo {title} {Angular momentum of the
  quantized electromagnetic field with periodic boundary conditions},\ }\href
  {https://doi.org/10.1103/PhysRevA.26.3428} {\bibfield  {journal} {\bibinfo
  {journal} {Phys. Rev. A}\ }\textbf {\bibinfo {volume} {26}},\ \bibinfo
  {pages} {3428} (\bibinfo {year} {1982})}\BibitemShut {NoStop}%
\bibitem [{\citenamefont {van Enk}\ and\ \citenamefont
  {Nienhuis}(1994)}]{Enk1994}%
  \BibitemOpen
  \bibfield  {author} {\bibinfo {author} {\bibfnamefont {S.~J.}\ \bibnamefont
  {van Enk}}\ and\ \bibinfo {author} {\bibfnamefont {G.}~\bibnamefont
  {Nienhuis}},\ }\bibfield  {title} {\bibinfo {title} {{Spin and Orbital
  Angular Momentum of Photons}},\ }\href
  {https://doi.org/10.1209/0295-5075/25/7/004} {\bibfield  {journal} {\bibinfo
  {journal} {{EPL}}\ }\textbf {\bibinfo {volume} {25}},\ \bibinfo {pages} {497}
  (\bibinfo {year} {1994})}\BibitemShut {NoStop}%
\bibitem [{\citenamefont {Holstein}\ and\ \citenamefont
  {Primakoff}(1940)}]{Holstein40}%
  \BibitemOpen
  \bibfield  {author} {\bibinfo {author} {\bibfnamefont {T.}~\bibnamefont
  {Holstein}}\ and\ \bibinfo {author} {\bibfnamefont {H.}~\bibnamefont
  {Primakoff}},\ }\bibfield  {title} {\bibinfo {title} {{Field Dependence of
  the Intrinsic Domain Magnetization of a Ferromagnet}},\ }\href
  {https://doi.org/10.1103/PhysRev.58.1098} {\bibfield  {journal} {\bibinfo
  {journal} {Phys. Rev.}\ }\textbf {\bibinfo {volume} {58}},\ \bibinfo {pages}
  {1098} (\bibinfo {year} {1940})}\BibitemShut {NoStop}%
\bibitem [{\citenamefont {Mitrofanov}\ and\ \citenamefont
  {Urazhdin}(2020)}]{Mitrofanov2020}%
  \BibitemOpen
  \bibfield  {author} {\bibinfo {author} {\bibfnamefont {A.}~\bibnamefont
  {Mitrofanov}}\ and\ \bibinfo {author} {\bibfnamefont {S.}~\bibnamefont
  {Urazhdin}},\ }\bibfield  {title} {\bibinfo {title} {Energy and momentum
  conservation in spin transfer},\ }\href
  {https://doi.org/10.1103/PhysRevB.102.184402} {\bibfield  {journal} {\bibinfo
   {journal} {Phys. Rev. B}\ }\textbf {\bibinfo {volume} {102}},\ \bibinfo
  {pages} {184402} (\bibinfo {year} {2020})}\BibitemShut {NoStop}%
\bibitem [{\citenamefont {Tramsen}\ \emph {et~al.}(2021)\citenamefont
  {Tramsen}, \citenamefont {Mitrofanov},\ and\ \citenamefont
  {Urazhdin}}]{Tramsen2021}%
  \BibitemOpen
  \bibfield  {author} {\bibinfo {author} {\bibfnamefont {N.}~\bibnamefont
  {Tramsen}}, \bibinfo {author} {\bibfnamefont {A.}~\bibnamefont
  {Mitrofanov}},\ and\ \bibinfo {author} {\bibfnamefont {S.}~\bibnamefont
  {Urazhdin}},\ }\href@noop {} {\bibinfo {title} {Effects of the dynamical
  magnetization state on spin transfer}} (\bibinfo {year} {2021}),\ \Eprint
  {https://arxiv.org/abs/arXiv:2101.08868} {arXiv:2101.08868} \BibitemShut
  {NoStop}%
\bibitem [{\citenamefont {Gol'dshtein}\ and\ \citenamefont
  {Tsukernik}(1984)}]{Tsukernik1984}%
  \BibitemOpen
  \bibfield  {author} {\bibinfo {author} {\bibfnamefont {E.~V.}\ \bibnamefont
  {Gol'dshtein}}\ and\ \bibinfo {author} {\bibfnamefont {V.~M.}\ \bibnamefont
  {Tsukernik}},\ }\bibfield  {title} {\bibinfo {title} {{Angular momentum of a
  Heisenberg ferromagnet with a magnetic dipole interaction}},\ }\href
  {http://www.jetp.ac.ru/cgi-bin/e/index/e/60/4/p764?a=list} {\bibfield
  {journal} {\bibinfo  {journal} {Sov. Phys. JETP}\ }\textbf {\bibinfo {volume}
  {60}},\ \bibinfo {pages} {764} (\bibinfo {year} {1984})}\BibitemShut
  {NoStop}%
\bibitem [{\citenamefont {Yan}\ \emph {et~al.}(2013)\citenamefont {Yan},
  \citenamefont {Kamra}, \citenamefont {Cao},\ and\ \citenamefont
  {Bauer}}]{Yan2013}%
  \BibitemOpen
  \bibfield  {author} {\bibinfo {author} {\bibfnamefont {P.}~\bibnamefont
  {Yan}}, \bibinfo {author} {\bibfnamefont {A.}~\bibnamefont {Kamra}}, \bibinfo
  {author} {\bibfnamefont {Y.}~\bibnamefont {Cao}},\ and\ \bibinfo {author}
  {\bibfnamefont {G.~E.~W.}\ \bibnamefont {Bauer}},\ }\bibfield  {title}
  {\bibinfo {title} {Angular and linear momentum of excited ferromagnets},\
  }\href {https://doi.org/10.1103/PhysRevB.88.144413} {\bibfield  {journal}
  {\bibinfo  {journal} {Phys. Rev. B}\ }\textbf {\bibinfo {volume} {88}},\
  \bibinfo {pages} {144413} (\bibinfo {year} {2013})}\BibitemShut {NoStop}%
\bibitem [{\citenamefont {Papanicolaou}\ and\ \citenamefont
  {Tomaras}(1991)}]{Papanicolaou1991}%
  \BibitemOpen
  \bibfield  {author} {\bibinfo {author} {\bibfnamefont {N.}~\bibnamefont
  {Papanicolaou}}\ and\ \bibinfo {author} {\bibfnamefont {T.}~\bibnamefont
  {Tomaras}},\ }\bibfield  {title} {\bibinfo {title} {Dynamics of magnetic
  vortices},\ }\href
  {https://doi.org/https://doi.org/10.1016/0550-3213(91)90410-Y} {\bibfield
  {journal} {\bibinfo  {journal} {Nucl. Phys. B}\ }\textbf {\bibinfo {volume}
  {360}},\ \bibinfo {pages} {425 } (\bibinfo {year} {1991})}\BibitemShut
  {NoStop}%
\bibitem [{\citenamefont {Tchernyshyov}(2015)}]{Tchernyshyov2015}%
  \BibitemOpen
  \bibfield  {author} {\bibinfo {author} {\bibfnamefont {O.}~\bibnamefont
  {Tchernyshyov}},\ }\bibfield  {title} {\bibinfo {title} {Conserved momenta of
  a ferromagnetic soliton},\ }\href
  {https://doi.org/https://doi.org/10.1016/j.aop.2015.09.004} {\bibfield
  {journal} {\bibinfo  {journal} {Ann. Phys. (N. Y.)}\ }\textbf {\bibinfo
  {volume} {363}},\ \bibinfo {pages} {98 } (\bibinfo {year}
  {2015})}\BibitemShut {NoStop}%
\bibitem [{\citenamefont {Dasgupta}\ and\ \citenamefont
  {Tchernyshyov}(2018)}]{Dasgupta2018}%
  \BibitemOpen
  \bibfield  {author} {\bibinfo {author} {\bibfnamefont {S.}~\bibnamefont
  {Dasgupta}}\ and\ \bibinfo {author} {\bibfnamefont {O.}~\bibnamefont
  {Tchernyshyov}},\ }\bibfield  {title} {\bibinfo {title} {Energy-momentum
  tensor of a ferromagnet},\ }\href
  {https://doi.org/10.1103/PhysRevB.98.224401} {\bibfield  {journal} {\bibinfo
  {journal} {Phys. Rev. B}\ }\textbf {\bibinfo {volume} {98}},\ \bibinfo
  {pages} {224401} (\bibinfo {year} {2018})}\BibitemShut {NoStop}%
\bibitem [{\citenamefont {Juraschek}\ and\ \citenamefont
  {Spaldin}(2019)}]{Jurascheck2019}%
  \BibitemOpen
  \bibfield  {author} {\bibinfo {author} {\bibfnamefont {D.~M.}\ \bibnamefont
  {Juraschek}}\ and\ \bibinfo {author} {\bibfnamefont {N.~A.}\ \bibnamefont
  {Spaldin}},\ }\bibfield  {title} {\bibinfo {title} {Orbital magnetic moments
  of phonons},\ }\href {https://doi.org/10.1103/PhysRevMaterials.3.064405}
  {\bibfield  {journal} {\bibinfo  {journal} {Phys. Rev. Materials}\ }\textbf
  {\bibinfo {volume} {3}},\ \bibinfo {pages} {064405} (\bibinfo {year}
  {2019})}\BibitemShut {NoStop}%
\bibitem [{\citenamefont {Neumann}\ \emph {et~al.}(2020)\citenamefont
  {Neumann}, \citenamefont {Mook}, \citenamefont {Henk},\ and\ \citenamefont
  {Mertig}}]{Neumann2020}%
  \BibitemOpen
  \bibfield  {author} {\bibinfo {author} {\bibfnamefont {R.~R.}\ \bibnamefont
  {Neumann}}, \bibinfo {author} {\bibfnamefont {A.}~\bibnamefont {Mook}},
  \bibinfo {author} {\bibfnamefont {J.}~\bibnamefont {Henk}},\ and\ \bibinfo
  {author} {\bibfnamefont {I.}~\bibnamefont {Mertig}},\ }\bibfield  {title}
  {\bibinfo {title} {Orbital magnetic moment of magnons},\ }\href
  {https://doi.org/10.1103/PhysRevLett.125.117209} {\bibfield  {journal}
  {\bibinfo  {journal} {Phys. Rev. Lett.}\ }\textbf {\bibinfo {volume} {125}},\
  \bibinfo {pages} {117209} (\bibinfo {year} {2020})}\BibitemShut {NoStop}%
\bibitem [{\citenamefont {Mecklenburg}\ and\ \citenamefont
  {Regan}(2011)}]{Mecklenburg2011}%
  \BibitemOpen
  \bibfield  {author} {\bibinfo {author} {\bibfnamefont {M.}~\bibnamefont
  {Mecklenburg}}\ and\ \bibinfo {author} {\bibfnamefont {B.~C.}\ \bibnamefont
  {Regan}},\ }\bibfield  {title} {\bibinfo {title} {Spin and the honeycomb
  lattice: Lessons from graphene},\ }\href
  {https://doi.org/10.1103/PhysRevLett.106.116803} {\bibfield  {journal}
  {\bibinfo  {journal} {Phys. Rev. Lett.}\ }\textbf {\bibinfo {volume} {106}},\
  \bibinfo {pages} {116803} (\bibinfo {year} {2011})}\BibitemShut {NoStop}%
\bibitem [{\citenamefont {Song}\ \emph {et~al.}(2015)\citenamefont {Song},
  \citenamefont {Paltoglou}, \citenamefont {Liu}, \citenamefont {Zhu},
  \citenamefont {Gallardo}, \citenamefont {Tang}, \citenamefont {Xu},
  \citenamefont {Ablowitz}, \citenamefont {Efremidis},\ and\ \citenamefont
  {Chen}}]{Song2015}%
  \BibitemOpen
  \bibfield  {author} {\bibinfo {author} {\bibfnamefont {D.}~\bibnamefont
  {Song}}, \bibinfo {author} {\bibfnamefont {V.}~\bibnamefont {Paltoglou}},
  \bibinfo {author} {\bibfnamefont {S.}~\bibnamefont {Liu}}, \bibinfo {author}
  {\bibfnamefont {Y.}~\bibnamefont {Zhu}}, \bibinfo {author} {\bibfnamefont
  {D.}~\bibnamefont {Gallardo}}, \bibinfo {author} {\bibfnamefont
  {L.}~\bibnamefont {Tang}}, \bibinfo {author} {\bibfnamefont {J.}~\bibnamefont
  {Xu}}, \bibinfo {author} {\bibfnamefont {M.}~\bibnamefont {Ablowitz}},
  \bibinfo {author} {\bibfnamefont {N.~K.}\ \bibnamefont {Efremidis}},\ and\
  \bibinfo {author} {\bibfnamefont {Z.}~\bibnamefont {Chen}},\ }\bibfield
  {title} {\bibinfo {title} {Unveiling pseudospin and angular momentum in
  photonic graphene},\ }\href {https://doi.org/10.1038/ncomms7272} {\bibfield
  {journal} {\bibinfo  {journal} {Nat. Commun.}\ }\textbf {\bibinfo {volume}
  {6}},\ \bibinfo {pages} {6272} (\bibinfo {year} {2015})}\BibitemShut
  {NoStop}%
\end{thebibliography}
\end{document}